# Revealing the role of interfacial heterogeneous nucleation in metastable thin film growth of rare-earth nickelates electronic transition materials

*Fengbo Yan[1][†], Zhishan Mi[1,2][†], Jinhao Chen[1], Haiyang Hu[1], Lei Gao[1,3]\*, Jiaou Wang[4]\*, Nuofu Chen[5], Yong Jiang[1], Lijie Qiao[1,3] and Jikun Chen[1]\**

[1]Beijing Advanced Innovation Center for Materials Genome Engineering, School of Materials Science and Engineering, University of Science and Technology Beijing, Beijing 100083, China

[2]China Iron & Steel Research Institute Group, Material Digital R&D Center, Beijing 100081, China

[3]Corrosion and Protection Center, University of Science and Technology Beijing, Beijing 100083, China

[4]Beijing Synchrotron Radiation Facility, Institute of High Energy Physics, Chinese Academy of Sciences, Beijing 100049, China

[5]School of Renewable Energy, North China Electric Power University, Beijing 102206, China

Correspondence: Prof. Lei Gao (gaolei@ustb.edu.cn), Prof. Jiaou Wang (wangjo@ihep.ac.cn), and Prof. Jikun Chen (jikunchen@ustb.edu.cn).

[†]Fengbo Yan and Zhishan Mi contributed equally to this work.

**Abstract:**

Although rare-earth nickelates (*Re*NiO$_3$, *Re*≠La) exhibit abundant electronic phases and widely adjustable metal to insulator electronic transition properties, their practical electronic applications are largely impeded by their intrinsic meta-stability. Apart from elevating oxygen reaction pressures, heterogeneous nucleation is expected as an alternative strategy that enables the crystallization of *Re*NiO$_3$ at low meta-stability. In this work, the respective roles of high oxygen pressure and heterogeneous interface in triggering *Re*NiO$_3$ thin films growth at metastable state are revealed. The *Re*NiO$_3$ (*Re*=Nd, Sm, Eu, Gd, and Dy) heterogeneous thin films growth on LaAlO$_3$ single crystal substrate have an effective crystallization at atmosphere without the necessity to apply high oxygen pressures, suggesting the interfacial bonding between the *Re*NiO$_3$ and substrates can sufficiently reduce the positive Gibbs formation energy of *Re*NiO$_3$, which is further verified by the first-principles calculations. Nevertheless, the abrupt electronic transitions only appear in *Re*NiO$_3$ thin films grown at high oxygen pressures, in which cases the oxygen vacancies are effectively eliminated via high oxygen pressure reactions as indicated by near-edge X-ray absorption fine structure (NEXAFS). This work unveils the synergistic effects of heterogeneous nucleation and high oxygen pressure on the growth of high quality *Re*NiO$_3$ thin films.

1. **Introduction**

The *d*-orbital correlation within transitional metal oxides enriches distinguished transportation functionalities such as metal to insulator transition (MIT) beyond conventional materials. One of the most important material family among the electronic transition oxide semiconductors is the rare-earth nickelates (*Re*NiO$_3$), the electronic orbital configuration of which experiences an abrupt and reversible transition from Ni$^{3\pm\Delta}$ to Ni$^{3+}$ via elevating temperature across a critical point ($T_{MIT}$). As a result, the energy band gap of *Re*NiO$_3$ merges abruptly with the electronic conductivity transforming from insulator (or semiconductor) to metal. One distinguished advantage of *Re*NiO$_3$ family is the adjustable $T_{MIT}$ with a broad range of 100 K-600 K by simply regulating their rare-earth composition [1, 2]. Apart from critical temperature, the MIT behavior of *Re*NiO$_3$ can be also triggered by stimuli such as polarized electric fields [3-5], mechanical or interfacial strains [6-8], and hydrogenation [9-14]. It is especially worthy to note that upon hydrogenation, the electronic structure of *Re*NiO$_3$ can be further reversibly triggered into the electron localized Ni$^{2+}$ configuration, and the hydrogenated *Re*NiO$_3$ is electron insulating but proton conducting. Owing to their extreme complexity in the electronic phase diagram and multiple electronic orbital transitions, the *Re*NiO$_3$ is demonstrated to be useful in correlated electronic devices such as mutant and NTCR thermistor [15], ocean electric field sensor [13], photocatalysis [16], correlated fuel cell [12], non-volatile logical devices [4], and correlated bio-sensor [17].

However, the above abundant applications in correlated electronics are largely bottlenecked by the intrinsic high metastability of *Re*NiO$_3$ (*Re*≠La). In contrast to conventional oxide semiconductors, *Re*NiO$_3$ (*Re*≠La) usually exhibits a positive formation free energy ($\Delta G$) [1, 18] and cannot be synthesized by conventional solid-state reactions. Previous reports indicated that the synthesis of metastable *Re*NiO$_3$ (*Re*≠La) relies on special conditions such as high oxygen pressure annealing [19, 20] and the template effect [6, 8, 21] from the substrate. For example, Ramanathan *et al*. [18] reported the non-epitaxial thin film growth of SmNiO$_3$ on silicon substrate via magnetron sputtering followed by high oxygen pressure annealing. They found that the non-epitaxial thin film growth of SmNiO$_3$ can be deposited successfully at the condition that $\Delta G$ is negative when the oxygen pressure and temperature are specific. Alternatively, the thin films of *Re*NiO$_3$ can be also epitaxially deposited on single crystalline perovskite substrates (e.g., LaAlO$_3$, SrTiO$_3$) by pulsed laser deposition (PLD) [22] and metal organic chemical vapor deposition (MOCVD) [23] without oxygen post annealing. Since the required oxygen pressure in PLD or MOCVD is far below the atmosphere, it unveils the importance of heterogeneous nucleation in the synthesis of rare-earth nickelates. Although *Re*NiO$_3$ films can be deposited by the two mentioned methods, vacuum combined with heterogeneous nucleation is merely used to grow the low and medium metastability *Re*NiO$_3$ films such as PrNiO$_3$, NdNiO$_3$, SmNiO$_3$, EuNiO$_3$ and GdNiO$_3$ and rarely reported to synthesize high metastability *Re*NiO$_3$ films (e. g. DyNiO$_3$ and ErNiO$_3$). In addition, the synergistic application of heterogeneous nucleation and high oxygen pressure annealing in spin coating-assisted epitaxy growth is more likely to achieve deposition of medium and high metastability $Re$NiO$_3$ films. Thus, it is still unclear that how the two growth conditions influence the thermodynamics of the growth of metastable $Re$NiO$_3$ film.

In this work, we unveil that the importance of interfacial heterogeneous nucleation via epitaxy to the material growth of metastable $Re$NiO$_3$. ($Re$=Sm, Nd, Eu, Gd and Dy). The respective mechanisms to form a distorted perovskite structure of $Re$NiO$_3$, as associated to the interfacial heterogeneous nucleation or the high oxygen pressure related solid state reactions, were separately investigated. The lattice constant and electronic transportation properties are systematically investigated for $Re$NiO$_3$ epitaxy on LaAlO$_3$ (001) and SrTiO$_3$ (001) substrates at various oxygen pressures from 1atm to 15 MPa and covering various meta-stabilities of $Re$NiO$_3$. Assisted by the near-edge X-ray absorption fine structure analysis (NEXAFS), the variation in electronic orbital structures associated to Ni-L edge and O-K edge were compared for $Re$NiO$_3$ grown at 1 atm and 15 MPa oxygen pressures. Apart from experimental investigations, density functional theory (DFT) calculations were performed to reveal the driving force of heterogeneous nucleation ascribed to chemical interface interaction between $Re$NiO$_3$ films and substrates. To clarify the dominant role to the formation of $Re$NiO$_3$, the reductions in free energy associated to the interfacial heterogeneous nucleation via epitaxy were calculated and further compared to the one associated to the elevation in oxygen pressure.

2. **Methods**

   **DFT calculations:** The DFT calculations were performed in the Vienna ab initio Simulation Package (VASP) [24]. The core electrons were described using the projector-augmented-wave (PAW) method and the electron exchange and correlation were modeled in GGA framework with the Perdewe-Burkee-Ernzerhof (PBE) form [25, 26]. All the calculations were carried out as spin-polarized. The plane-wave cutoff energy was set to 400 eV. During the geometrical optimizations, atoms were allowed to be relaxed until the forces on all the relaxed atoms were less than 0.01 eV/Å.

   **Sample preparation:** The $Re$NiO$_3$ ($Re$ = Sm, Nd, Eu, Gd and Dy) thin films were grown via the wet chemical method using spin coating. $Re$(NiO$_3$)$_3$·6H$_2$O and Ni(OCOCH$_3$)$_2$·4H$_2$O were dissolved in ethylene glycol monomethyl ether according to the molar stoichiometric ration 1:1 and stirred for 2h using the magnetic stirrer. The obtained chemical solution was deposited on the LaAlO$_3$ (LAO) (001) and SrTiO$_3$ (STO) (001) substrates respectively by spin coating and then the substrates were dried at 175℃ for 90s. Subsequently, the as-deposited thin films were annealed in a tubular furnace (OTF-1200X-HP-30) at 800°C for 3h with the heating rate 10°C/min to form single crystal $Re$NiO$_3$ thin films under different oxygen pressure.

   **Characterizations:** These samples were characterized by X-ray diffraction (XRD) for phase identifications and analysis of variation of lattice constants. The resistivity of thin films was measured in vacuum within the temperature range from 300K to 500K using a commercialized CTA-system and from 5K to 300K using physical property measurement system (PPMS, Quantum Design Inc.), respectively. In order to reflect the valence variations in electronic structures of $Re$NiO$_3$, near-edge X-ray absorption fine structure (NEXAFS) was performed at BSRF-4b9b beam line of Beijing Synchrotron Radiation Facility, Institute of High Energy Physics, Chinese Academy of Sciences.

## 3. Results and discussion

*3.1. Role of interfacial nucleation via epitaxy on the lattice template of the substrate*

The *Re*NiO$_3$ usually exhibits positive Gibbs free energy (Δ$G$) under the conventional growth conditions [1, 18], and therefore the *Re*NiO$_3$ cannot be synthesized using conventional solid-state reactions. Nevertheless, the *Re*NiO$_3$ thin films such as SmNiO$_3$, NdNiO$_3$ and PrNiO$_3$, can be indeed epitaxially grown on single crystalline perovskite substrates utilizing the template effects of the substrate lattice that triggers heterogeneous nucleation [8, 27, 28]. In order to investigate the nucleation and growth mechanism of *Re*NiO$_3$ films associated to such heterogeneous process, a series of experimental conditions were used to grow *Re*NiO$_3$ films, including SmNiO$_3$(SNO), NdNiO$_3$(NNO), EuNiO$_3$(ENO) and GdNiO$_3$(GNO) on perovskite substrates via spin coating followed by high oxygen pressure annealing under various pressure.

In our experiment, *Re*NiO$_3$ (*Re* = Sm, Nd, Eu and Gd) films were grown on LAO (001) and STO (001) substrates by the wet chemical method using spin coating then annealed at 800 °C under 1 atm and 15MPa oxygen pressure. It is noteworthy that *Re*NiO$_3$ films with medium and high metastability (e. g. EuNiO$_3$) cannot be synthesized on STO (001) substrates successfully [23], as is indicated by no resistance for these *Re*NiO$_3$ films. Previous works have reported that there exists a large lattice mismatch between *Re*NiO$_3$ and STO substrates in particular with *Re* element after Eu [29], leading to the formation of relaxed array of nanocrystalline islands in the *Re*NiO$_3$ films [22], which has also been observed in other perovskite oxide films such as PbTiO$_3$/SrTiO$_3$ [30]. Therefore, it is prone to form misfit dislocations or structure defects throughout the film especially in the low oxygen pressure, for example, BaTiO$_3$/SrTiO$_3$ [31]. Since the films cannot be integrated together, the sample resistances are usually beyond the range of multimeter. In addition, it was also affirmed by reciprocal space mapping that the interfacial strain between *Re*NiO$_3$ films and STO substrate was relaxed [7]. From these aspects, the experimental results from various *Re*NiO$_3$ thin films deposited on LAO substrates are more reliable and convincing compared to those from *Re*NiO$_3$ thin films on STO substrates.

To characterize structure quality of samples grown via heterogeneous nucleation followed by oxygen annealing process, X-ray diffraction (XRD) was performed and the results are demonstrated in Fig.1. Fig. 1a-c shows the X-ray *θ-2θ* scans taken on *Re*NiO$_3$ (*Re* = Sm, Nd and Eu) films grown on LAO (001) substrates under the oxygen annealing pressure of 1atm and 15 MPa at 800°C. The diffraction peaks of SNO, NNO and ENO thin films locate beside the diffraction peaks of LAO substrates, indicating the single crystalline of these thin films grown on LAO (001) substrates at both oxygen annealing pressure of 1 atm and 15 MPa (see XRD patterns at a broader scanning range in Fig. S1. While some additional diffraction peaks can be seen in these XRD patterns, the intensity of these additional diffraction peaks is much lower than that of the (00*l*) peaks associated with *Re*NiO3 thin films such as (001), (002) and (003). The advantageous interfacial strain between as-grown *Re*NiO3 thin films and the LAO substrates has been reported in our previous reports [7, 32, 33]. Since the *Re*NiO3 films can be grown on LAO substrates even under low oxygen pressure, the heterogeneous nucleation rather than oxygen pressure plays an important role in stabilizing the distorted perovskite structure of as-grown *Re*NiO3 in metastable manner. In the past, it was generally accepted that strain energy between *Re*NiO3 thin film and substrate plays a crucial role in the nucleation stage of rare earth nickelate phase because *Re*NiO3 phase is only deposited on perovskite substrates such as LaAlO3 and SrTiO3 while only the mixture of rare-earth oxide and NiO phase is grown on other type substrates such as MgO, YSZ [34] and ZrO2(Y2O3) [23]. Besides, the magnitude of the strain also has an important influence on the formation of *Re*NiO3 phase. The smaller lattice strains of *Re*NiO3 thin films on LAO than STO substrates shown in Table 1 confirm the growth of *Re*NiO3 films on LAO is favorable. For example, GdNiO3 single phase is only successfully deposited on LAO substrate owing to lower lattice mismatch [23, 35]. However, the thickness of samples prepared in our experiment is about 10nm and therefore the strain energies between *Re*NiO3 thin films and LAO substrates are expected to be not quite large when the volume of nucleus is small enough. In this case, strain energy will be not sufficient to overcome the nucleation barrier, instead interfacial chemical energy will become a key role to decide whether heterogeneous nucleation is formed. In order to verify this point, the DFT calculations of the interface free energies of samples were performed, as is shown in Fig. 2. Fig. 2a-b illustrate the crystal structures of *Re*NiO3 and substrates for LAO and STO, respectively. Fig. 2c shows SmNiO3/LaAlO3 and SmNiO3/SrTiO3 interface structures adopted in DFT calculations. Fig. 2d shows the interface free energies of as-grown *Re*NiO3 films on LAO and STO substrates, where the specific interface free energy is defined as:

$$E_i = (E_{interface} - E_{up} - E_{down})/S \tag{1}$$

where $E_{interface}$ is the total interface free energy of the interface structure. $E_{up}$ is the energy of the upper *Re*NiO3 film. $E_{down}$ is the energy of the lower substrate and *S* is the area of interface. All the calculated interface free energies are negative and therefore the interface free energy is expected to descend the pristine positive free energy of *Re*NiO3 to trigger heterogeneous nucleation of *Re*NiO3 films on LAO and STO substrates. In addition, the more negative interface free energies when epitaxy *Re*NiO3 thin films grown on LAO compared to STO indicate the interfaces of *Re*NiO3 films on LAO are more stable. The interfacial charge transfer of the interface structure can further describe the interfacial bonding strength between *Re*NiO3 films and LAO and STO substrates, and the calculated results are shown in Fig. 2e-f. The differential charge density at the interface along the c-axis direction is defined as

$$\Delta\rho = \rho_{interface} - \rho_{up} - \rho_{down} \tag{2}$$

where $\rho_{interface\ structure}$ represents the total charge density of the system, $\rho_{up}$ and $\rho_{down}$ represent the charge density of upper film structure and lower substrate structure. A positive value of $\Delta\rho$ represents valence electron increasing and a negative value of $\Delta\rho$ represents electron decreasing. The *Re*NiO3 films grown on LaAlO3 have larger charge transfer magnitudes comparing with those on SrTiO3 (as shown in Fig. S2, indicating stronger interface bonding strength and larger film growth driven force. Thus, smaller interface lattice strain and the more negative interface free interfaces are the dominant factors to facilitate the heterogeneous nucleation of *Re*NiO3.

**Table 1**
Lattice strain of the a-axis and b-axis (More details are shown in section 2 of supplementary material). The results indicate the oxides grown on LaAlO3 have smaller lattice strain.

|                     | LaAlO3  |        | SrTiO3  |        |
|---------------------|---------|--------|---------|--------|
|                     | a       | b      | a       | b      |
| **SmNiO3**          | -0.021  | 0.008  | -0.056  | -0.028 |
| **NdNiO3**          | -0.014  | 0.013  | -0.049  | -0.024 |
| **EuNiO3**          | -0.011  | -0.009 | -0.047  | -0.045 |
| **GdNiO3**          | -0.029  | 0.021  | -0.064  | -0.016 |

*3.2. Role of the elevation in oxygen pressure*

Apart from heterogeneous nucleation, oxygen pressure was also expected as another key point to influence the growth of *Re*NiO3 films [36]. To further investigate the properties of metastable *Re*NiO3 associated to different oxygen pressures, the electrical transportation properties of *Re*NiO3 films under two selected oxygen pressure conditions are investigated. The measured temperature-dependent in-plane electrical resistivity for *Re*NiO3 thin films with various rare-earth elements (*Re* = Sm, Nd, Eu and Gd) on LAO and STO substrates at 1atm and 15 MPa oxygen pressure is shown in Fig. 3. The MIT behaviors of the as-deposited SNO/LAO and SNO/STO under 1atm measured via both heating up and cooling down processes are shown in Fig. 3a. It is worth noting that for SNO/LAO and SNO/STO under 1atm, the obtained MIT behavior is very flatten compared with those as-deposited at 15 MPa oxide pressure, indicating the MIT behavior is suppressed for the *Re*NiO3 film grown under 1 atm. As for the $\rho$-$T$ curves, it can be observed that the SNO/LAO system heating up and cooling down profiles are sharper than those of SNO/STO system, which is attributed to large interface strain of the SNO/STO system [7]. Meanwhile, as shown in Fig. 3b, the $\rho_T/\rho_{300K}$ decreases more remarkably for the SNO/LAO system than that of SNO/STO system at 15 MPa oxygen pressure, which is consistent with the results in Fig. 3 a. In Fig. 3c-d, the MIT behaviors for NNO, ENO and GNO on LAO under 1atm and 15 MPa oxygen pressure are similar compared to SNO. However, it is also noteworthy that an obvious hysteresis in $T_{MIT}$ can be seen for NNO on LAO and STO as shown in Fig. 3c when measuring the *R-T* curve via heating up compared to cooling down, which has also been reported by the previous literatures [7, 37, 38] that the $\rho$-$T$ curves measured through heating up and cooling down exist a hysteresis for *Re*NiO3 with larger rare-earth elements taking up the A-site of the perovskite structure.

Therefore, it can be found that without the high oxygen pressures, the MIT behaviors of $Re$NiO$_3$ films on LAO and STO substrates under 1 atm behave poorly, which might be caused by oxygen vacancies since $Re$NiO$_3$ are prone to form oxygen vacancies under low oxygen pressure [36, 39]. It has been reported that the presence of oxygen vacancies can alter the Ni valence state from Ni$^{3+}$ to Ni$^{2+}$ via offering electrons from the missing oxygen atoms [36, 40-42]. Moreover, owing to radius of Ni$^{3+}$ (r=0.56Å) being smaller than that of Ni$^{2+}$(r=0.69Å), oxygen vacancies enlarge out-of-plane lattice parameter, which is verified by Fig. 1d. As for in-plane lattice, owing to the good lattice match between LAO and $Re$NiO$_3$, the in-plane lattice is independent of the oxygen pressure and coincide with that of LAO. In this case, it can be further inferred that the expansion of the lattice constants of $Re$NiO$_3$ expand Ni-O bonds and changes the rotations and tilts of NiO$_6$ octahedra [43]. Further, the larger Ni-O bonds can also induce the smaller Ni-O-Ni bond angle owing to the reduce of tolerance factor, which decreases the bandwidths. Moreover, the reduced bandwidths make the Coulomb repulsion between the electrons stronger, as a result, leading to the suppression of MIT behaviors of $Re$NiO$_3$ films grown under low oxygen pressure. More importantly, the high oxygen pressure provides sufficient oxygen atoms for the growth of $Re$NiO$_3$ films on substrates to ensure elements in $Re$NiO$_3$ satisfying the specific stoichiometric ratio.

In order to verify the oxygen vacancies indeed alter the valence state of Ni from +3 to +2 under low oxygen pressure, the near edge X-ray absorption fine structure (NEXAF) analyses have been performed and the results are shown in Fig. 4. Comparing the O: K-edge EXAFS spectra of SmNiO$_3$ and NdNiO$_3$ at 1atm with those at 15MPa oxygen pressure in Fig. 4a and 4c respectively, we can distinctly observe the increased pre-peak (the first peak in the low photon energy) in the spectra for SNO/LAO and NNO/LAO under the higher oxygen pressure. The O-K edge pre-peak is related to the strong covalency between the Ni 3d and O 2p states. Therefore, the low O-K edge pre-peak intensity suggests the large amount of oxygen vacancies which alter the Ni$^{3+}$ to Ni$^{2+}$. In this case, to some degree, the pre-peak in their O: K-edge spectra reflects the relative content of Ni$^{3+}$ or Ni$^{2+}$ [7], which means the higher content of Ni$^{3+}$ owing to the higher O-K edge pre-peak intensity when the annealing oxygen pressure is 15MPa for both SNO/LAO and NNO/LAO. With respect to the Ni: $L_3$ spectra, the split peaks in Fig. 4b and 4d originate from the transition from Ni-2p to Ni-3d, which suggests the proportion of the Ni$^{3+}$ compared to the Ni$^{2+}$ is represented by the relative height of peak in high photon energy to the peak in low photon energy for SNO/LAO and NNO/LAO, respectively. Therefore, the as-grown SNO thin film and NNO thin film on LAO substrates under 15MPa contain more Ni$^{3+}$, further confirming that oxygen vacancies change the valence state of Ni from +3 to +2. In addition, it can also be seen that under the same synthesis condition, the relative content of Ni$^{3+}$ in NNO/LAO is higher than that in SNO/LAO from the Ni: $L_3$ edge and O: $K$ edge spectra respectively which is consistent with the previous reports [7, 44, 45].

Meanwhile, to explicate the role of oxygen vacancies on the transport properties of $Re$NiO$_3$ in the insulator region, the low temperature electrical transportation of the $Re$NiO$_3$/LaAlO$_3$ (001) were characterized and analyzed. The evolutions of $\rho_T/\rho_{300K}$ versus temperature about $Re$NiO$_3$/LaAlO$_3$ systems are illustrated in Fig. 5a, which suggests the profiles of SNO, ENO and GNO under 15MPa are steeper compared to those under the low oxygen pressure. According to references [46-49], the $Re$NiO$_3$ generally exhibits a Mott variable range hopping (VRH) mechanism in the insulator region, which can be confirmed by the linearity of $\ln(\rho_T/\rho_{300K})$-$T^{-0.25}$. VRH model describes the transportation mechanism of carriers in the low temperature range that a carrier hops to the site with a closer energy rather than the nearest position [50]. It can be described by means of the following expression

$$\rho(T) = \rho_0 \exp\left(\frac{T_0}{T}\right)^p \tag{3}$$

where $\rho_0$ is the prefactor, $T_0$ is the characteristic temperature, $p$ is the exponent related to the conduction mechanism. In three dimensions, $p=1/4$ and $T_0$ can be determined by

$$T_0 \equiv T_{Mott} = \frac{18}{k_B N(E_F) a^3} \tag{4}$$

where $N(E_F)$ is the density of states (DOS) close to Fermi energy ($E_F$) and a is the localization length of carriers. More details of the fitting results of $\ln(\rho_T)$-$T^{-0.25}$ are given in Fig. S3 as discussing carrier transport model of $Re$NiO$_3$ in three dimensions.

In the meanwhile, the thickness of the $Re$NiO$_3$ grown by chemical depositions is approximately 10nm. Therefore, we also fitted the $\rho_T$-$T$ tendencies via two dimensional carrier transport model that can be expressed as $\ln(\rho_T)$-$T^{-1/3}$, as shown in Fig. S4. The values of $T_{Mott}$ obtained in three dimensions are illustrated in Fig. 5b, which are generally smaller in low oxygen pressure. According to the reference [51], $T_{Mott}$ characterizes the degree of the disorder. Fig. 5b indicates the disorder of $Re$NiO$_3$ grown on LAO via chemical deposition increases with the oxygen pressure increasing, since the deficient oxygen atoms in low oxygen pressure are apt to form oxygen vacancies in the process of $Re$NiO$_3$ films growth, leading to the formation of Ni$^{2+}$ and decrease of disorder. Thus, the oxygen pressure plays a critical role in controlling stoichiometric ratio of $Re$NiO$_3$. Meanwhile, it is worthy to be mentioned that the behavior of $T_{Mott}$ of other $Re$NiO$_3$ is consistent with SmNiO$_3$. Nevertheless, it has been reported [52] that $Re$NiO$_3$ is a multiple-band system with both electron- and hole-like carriers and therefore the variation of carrier mobility under low temperature is very complicate, possibly leading to the appearance of some anomalous points, such as EuNiO$_3$ annealed under 4Mpa and GdNiO$_3$ annealed at 1atm. In Fig. 5c, the seebeck coefficient of $Re$NiO$_3$ thin films on LAO substrates at room temperature is plotted as a function of the oxygen pressure. The obtained results show the seebeck coefficients at room temperature of $Re$NiO$_3$ thin films such as SmNiO$_3$, NdNiO$_3$ and EuNiO$_3$ keep negative whether annealed in low or high oxygen pressure, while the seebeck coefficient of GNO changes from negative to positive when oxygen pressure increases from 1atm to 15MPa, suggesting the carrier type becomes p-type. Similar phenomenon has also been reported in reference [53] and the reason is expected to be oxygen vacancies.

According to above results, it can be concluded that high oxygen pressure is beneficial to maintaining the valance state of $Ni^{3+}$ by eliminating oxygen vacancies and abrupt electronic transitions of $Re$NiO$_3$ films.

*3.3 Metastable ReNiO$_3$ thin film growth combining epitaxy and high oxygen pressure*

Since heterogeneous nucleation reduces the intrinsic positive Gibbs free energy and high oxygen pressure can eliminate oxygen vacancies to maintain the valance state of $Ni^{3+}$, the thin film growth of $Re$NiO$_3$ using chemical based spin coating process provides the opportunity to achieve thin film growth of $Re$NiO$_3$ with high metastability (e. g. $Re$ from Dy to Tm) that has been rarely reported previously [33, 44]. As a representative case, the DyNiO$_3$ can be epitaxially grown on LaAlO$_3$ (001) single crystalline substrate via spin coating process of Dy(NO$_3$)$_3$·6H$_2$O, Ni(OCOCH$_3$)$_2$·4H$_2$O precursors dissolved in 2-methoxyethanol followed by annealing at 15MPa oxygen pressure at 800℃. Fig. 6a shows the *R-T* tendency of as-grown DNO/LAO, where a remarkable negative temperature coefficient resistance (NTCR) property is observed although the resistivity of DyNiO$_3$ thin film on LAO substrate deviates the normal values when temperature reaches to 525K that is not MIT temperature. The deviation could be attributed to high metal-to-insulator transition temperature destroys the integrity of interfacial structure between DNO thin films and LAO substrates. And the corresponding crystal structure illustration of DyNiO$_3$ is shown in Fig. 6c. Although the complete *R-T* curve reflecting the metal to insulator transition is not measured successfully in DyNiO$_3$ thin films, the temperature of metal to insulator transition has rarely been obtained via resistivity variation as a function of temperature for high metastability $Re$NiO$_3$ powders such as DyNiO$_3$, HoNiO$_3$ in previous reports. For example, García-Muñoz *et al*. [54] provided *R* vs *T* curves for YNiO$_3$ ceramics up to 750K with a quite high oxygen pressure to synthesize. As a result, the determination of $T_{MIT}$ for high metastability $Re$NiO$_3$ powders generally utilizes the DSC analysis [55] and neutral diffraction [56].

Another overwhelming advantage to grow $Re$NiO$_3$ thin films using chemical based precursors, compared to the previously reported vacuum-based approaches, is the capability to achieve more flexible and convenient adjustment in the rare-earth composition. The rare-earth composition within $Re$NiO$_3$ regulates the distortion in NiO$_6$ octahedron that determines the electronic structure and $T_{MIT}$, and to achieve the continuous regulation of $T_{MIT}$ within a wide range of temperature, the effective growth of $Re$NiO$_3$ with multicomponent of $Re$ is required. Besides, the growth of $Re$NiO$_3$ thin films with various $Re$ elements can not only help us to know how to stabilize of such a multi $Re$ system in a single crystalline to further explore electronic behavior and metal to insulator transition mechanism but also allow us to investigate the phase transition mechanism around the triple point of the $Re$NiO$_3$ phase diagram [57, 58]. Patel *et al*. [57] has grown nickelate films with 5 different cations in the A-site and excellent crystallinity using PLD, the variation of $Re$ compositions in target preparation process is not convenient. Compared to the vacuum-based approaches such as PLD or magnetron sputtering, the rare-earth composition of the chemical deposited $Re$NiO$_3$ can be flexibly and quantitatively regulated by the combination of $Re$(NiO$_3$)$_3$·6H$_2$O with various compositions of $Re$ according to the required stoichiometric ratio. As a representative example, Nd$_{0.3}$Sm$_{0.4}$Gd$_{0.3}$NiO$_3$ thin film was grown on LAO substrates via spin coating method of Nd(NiO$_3$)$_3$·6H$_2$O, Sm(NiO$_3$)$_3$·6H$_2$O and Gd(NiO$_3$)$_3$·6H$_2$O and Ni(OCOCH$_3$)$_2$·4H$_2$O precursors dissolved in 2-methoxyethanol according to the molar ratio 0.3:0.4:0.3:1 followed by annealing at 15MPa oxygen pressure at 800℃. Fig. 6b shows the resistivity variation as a function of temperature and the metal to insulator transition temperature is estimated around 373K. Meanwhile, the corresponding crystal structure illustration for Nd$_{0.3}$Sm$_{0.4}$Gd$_{0.3}$NiO$_3$ is shown in Fig. 6d.

*3.4. Estimating the free energy reduction of interface chemical bonds from thermodynamic perspective*

In previous researches, it was generally known that strain induced by the film-substrate mismatch is the most factor to influence the nucleation of $Re$NiO$_3$ phase [22]. However, it is worthy of note that strain energy may not work when both the nucleation volume and thickness of the thin film are very small [35], suggesting another factor is critical in the formation of $Re$NiO$_3$ phase. It is expected to be the interface free energy.

In order to explain the role of interface free energy, we estimated the interface free energy and Gibbs free energy of formation for $Re$NiO$_3$ as the following: the interface free energy was calculated by DFT modeling while the Gibbs free energy of formation for bulk $Re$NiO$_3$ were obtained when $Re$NiO$_3$ thin films were grown on substrates with no lattice mismatch such as oxidized silicon wafers. The DFT calculated results are listed in the Table 2. And the formation of bulk $Re$NiO$_3$ in our experiment can be expressed as the following:

$$0.5Re_2O_3 + NiO + 0.25O_2 = ReNiO_3 \qquad (5)$$

where $Re$ is the rare-earth element.

According to the reference [18], the molar Gibbs free energy of formation for bulk $Re$NiO$_3$ via reaction (6) can be expressed as the following formula:

$$\Delta G = \Delta H_{LNO,1000k} - T\Delta S_{LNO,1000K} + (h - sT)\big(r(Re^{3+}) - r(La^{3+})\big) -$$
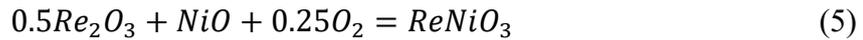
$$(1/4)RT\ln(P/p) \qquad (6)$$

LNO stands for LaNiO$_3$, $R$ is the ideal gas constant, the $P$ accounts for oxygen pressure with MPa unit, and $p$ represents the standard atmospheric pressure (0.1MPa). According to references [18, 59], the enthalpy and entropy changes for this reaction at 1000K are $\Delta H_{LNO,1000K}$=-46.07kJmol$^{-1}$ and $\Delta S_{LNO,1000K}$=-26.4×10$^{-3}$kJ(K$^{-1}$mol$^{-1}$). $h$ and $s$ obtained from reference [18] represent the trend in enthalpy and entropy respectively. The values of them are $h$ = -356kJmol$^{-1}$Å$^{-1}$, $s$=-0.125kJK$^{-1}$mol$^{-1}$Å$^{-1}$ (see more details in Section 5 of supplementary material). $r(Re^{3+})$ and $r(La^{3+})$ are the radii of $Re^{3+}$ and La$^{3+}$, respectively. Here, we use $r$(Sm$^{3+}$) =1.24 Å, $r$(Nd$^{3+}$) =1.27 Å, $r$(Eu$^{3+}$) =1.20 Å and $r$(Gd$^{3+}$) =1.19 Å, which have been reported in reference [60]. Consequently, the thermodynamic phase stability diagrams for SNO, NNO, ENO and GNO are shown in Fig. 7, along with points representing reported synthesis conditions for these materials.

As shown in Fig. 7, the ordinary synthesis conditions of the powder SNO and NNO are above the white region, which are indicated by solid cyan circle and solid orange diamond for SNO powder in Fig. 7a and solid cyan circle for NNO powder in Fig. 7b, respectively. However, the other points below the white region in Fig. 7 have been also confirmed to achieve the growth of $Re$NiO$_3$ films on LaAlO$_3$ substrates. It has been mentioned above that the main reason is heterogeneous nucleation stabilizing the distorted perovskite structure in metastable manner, whose driving force is practically from the interface free energies between $Re$NiO$_3$ films and substrates. To reveal this interface chemical effect, the $\Delta G$ calculated by eqn (7) for every condition to grow $Re$NiO$_3$ films on LAO substrates and the corresponding interfacial free energies calculated by DFT are listed and demonstrated in the Table 2.

To enable the comparison between the Gibbs free energy ($\Delta G$) obtained from eqn (6) and interfacial chemical energy calculated by DFT, we assumed the following conditions: (1) the film thickness is 10nm, (2) a molar density is approximately 0.015mol cm$^{-3}$ [18], (3) the interfacial free energy remains constant when the oxygen pressure varies. It is noteworthy in Table 2 that the Gibbs free energy of formation of bulk $Re$NiO$_3$ in reaction (5) is less than the interfacial chemical energy for SNO/LAO, NNO/LAO, ENO/LAO and GNO/LAO under the given conditions, which means the $Re$NiO$_3$ thin films ($Re$=Sm, Nd, Eu and Gd) can be deposited on LaAlO$_3$ substrates and further affirms the interface free energy is efficient to reduce the Gibbs free energy. These discoveries give us a new insight into homogeneous nucleation and growth of $Re$NiO$_3$. For example, it is commonly recognized that heterogeneous nucleation caused by substrate template effect is not as important as high oxygen pressure to determine the nucleation of $Re$NiO$_3$ phase. However, the experimental and DFT calculation results in this study suggest in spin coating-assisted epitaxy growth, heterogeneous nucleation dependent on substrate template effect is the key factor to induce the nucleus of $Re$NiO$_3$ while the role of high oxygen pressure is essential to eliminate oxygen vacancies and maintain the valance of Ni$^{3+}$ within $Re$NiO$_3$ to achieve an abrupt MIT performance.

**Table 2**
Gibbs free energy of formation of bulk $Re$NiO$_3$ according to eqn (6) for different synthesis conditions and the Interface free energy for $Re$NiO$_3$/LAO (001)

| samples | Condition | $\Delta G$(kJ/mol) | $\Delta G$(eV/Å$^2$) | Interface free energy(eV/Å$^2$) |
|---|---|---|---|---|
| SNO/LAO | 1273K 7MPa | -0.0790 | -7.4084e-04 | -0.23072 |
|  | 953K 10$^{-3}$MPa | 16.6362 | 0.1560 |  |
|  | 953K 0.1MPa | 7.5142 | 0.0704 |  |
|  | 1073K 15MPa | -2.2927 | -0.0215 |  |
| NNO/LAO | 953K 0.1MPa | 0.4080 | 0.0038 | -0.22188 |
|  | 1123K 0.1MPa | 2.9835 | 0.0280 |  |

|  | 1073K 15MPa | -8.9489 | -0.0839 |  |
| --- | --- | --- | --- | --- |
| ENO/LAO | 1073K 15MPa | 6.5823 | 0.0617 | -0.23848 |
| GNO/LAO | 1073K 15MPa | 8.8011 | 0.0825 | -0.23025 |

4. **Conclusions**

In conclusion, the dominant roles of interfacial heterogeneous nucleation and high oxygen pressure related solid state reactions on the crystallization and growth of $Re$NiO$_3$ ($Re$=Sm, Nd, Eu, Gd and Dy) covering various metastability are distinguished. From the one aspect, the interfacial template effect between $Re$NiO$_3$ and single crystalline perovskite oxide substrate triggers the heterogeneous nucleation and results in the nucleation of metastable $Re$NiO$_3$ without the necessity to apply high oxygen pressures. This understanding is well supported by the density functional theory (DFT) calculations, where the driving force of heterogeneous nucleation is ascribed to chemical interface interactions between films and substrates that sufficiently descend the positive formation energy of $Re$NiO$_3$. From the other aspect, although the metastable $Re$NiO$_3$ can be stabilized by epitaxy on LaAlO$_3$ substrate at normal pressure, their metal to insulator transition abruption is largely reduced owing to the derivation of electronic structures from Ni$^{3+}$ towards Ni$^{2+}$ as indicated by near-edge X-ray absorption fine structure analysis. This highlights the necessity of the high oxygen pressure in compensating the oxygen vacancy within $Re$NiO$_3$ that maintains the abruption of their temperature induced metal to insulator transitions. The synergistic effects of heterogeneous nucleation and high oxygen pressure guarantee the spin coating-assisted epitaxy growth of $Re$NiO$_3$ thin films with high intrinsic metastability.

**Acknowledgments**
This work was supported by the National Key Research and Development Program of China (No. 2018YFB0704300) and the National Natural Science Foundation of China (No. 62074014 and 52073090), the Fundamental Research Funds for the Central Universities (No. FRF-TP-19-023A3Z), and the Beijing New-star Plan of Science and Technology (No. Z191100001119071).

**Competing interests**
We declare no competing financial interest.

**Additional information:** Supplementary Information is available for this manuscript.

**Correspondence:** Prof. Lei Gao (gaolei@ustb.edu.cn), Prof. Jiaou Wang (wangjo@ihep.ac.cn), and Prof. Jikun Chen (jikunchen@ustb.edu.cn).

**Figure and Caption**

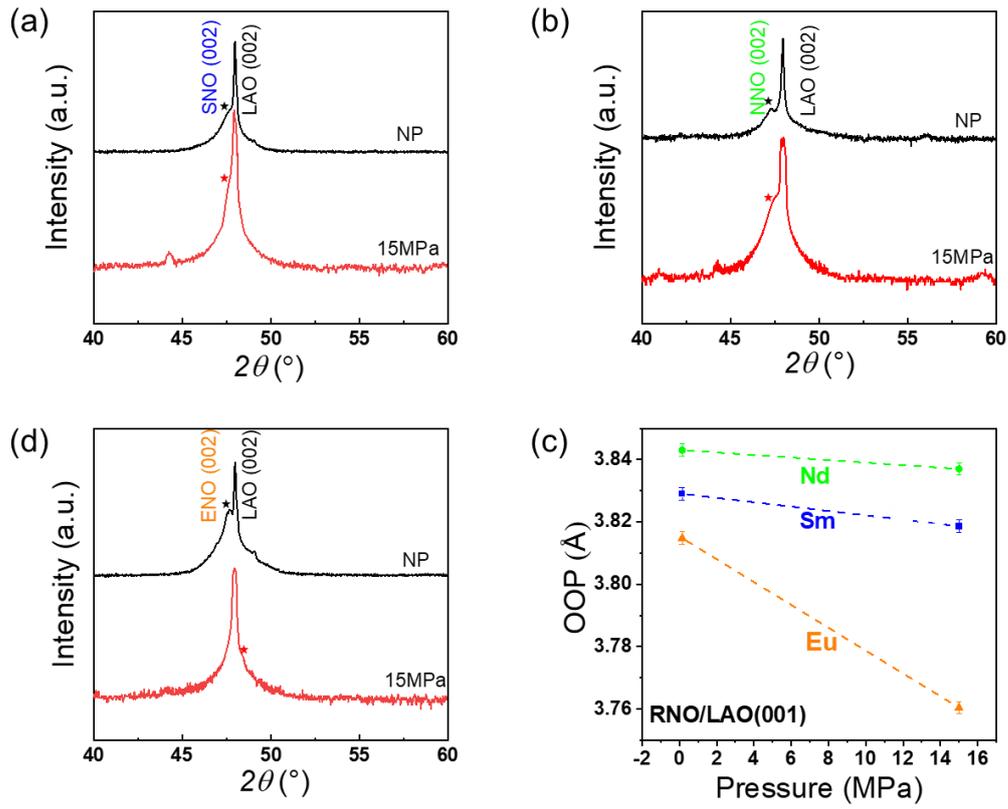

**Fig. 1.** X-ray diffraction patterns (*θ-2θ* scans) of as-grown (a) SmNiO$_3$/LaAlO$_3$ (001) (b) NdNiO$_3$/LaAlO$_3$ (001) and (c) EuNiO$_3$/LaAlO$_3$(001) under the annealing pressure of normal pressure and 15MPa with the same annealing temperature (*T*=800°C). (d) Calculated results of out-of-plane lattice constants for as-deposited *Re*NiO$_3$/LaALO$_3$(001) (*Re*=Sm, Nd, Eu) thin films. The XRD data for SNO/LAO and ENO/LAO under 15MPa are from the reference [61].

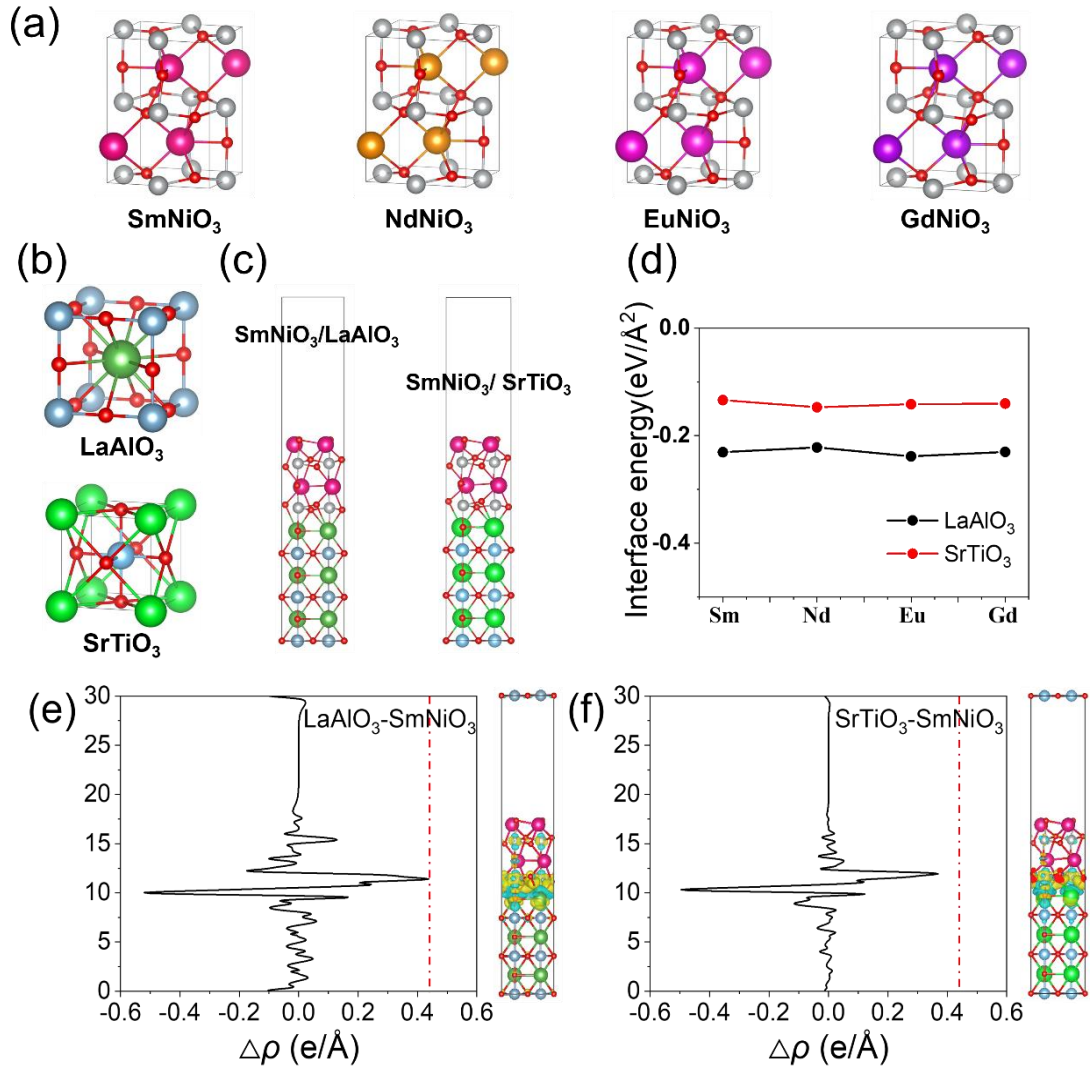

**Fig. 2.** (a) Unit cells of SmNiO$_3$, NdNiO$_3$, EuNiO$_3$, GdNiO$_3$. (b) Unit cells of LaAlO$_3$ and SrTiO$_3$. (c) SmNiO$_3$-LaAlO$_3$ interface and SmNiO$_3$-SrTiO$_3$ interface. (d) Interface energies of eight different interfaces. (e) the interfacial charge transfer of LaAlO$_3$-SmNiO$_3$ interface (f) SrTiO$_3$-SmNiO$_3$ interface.

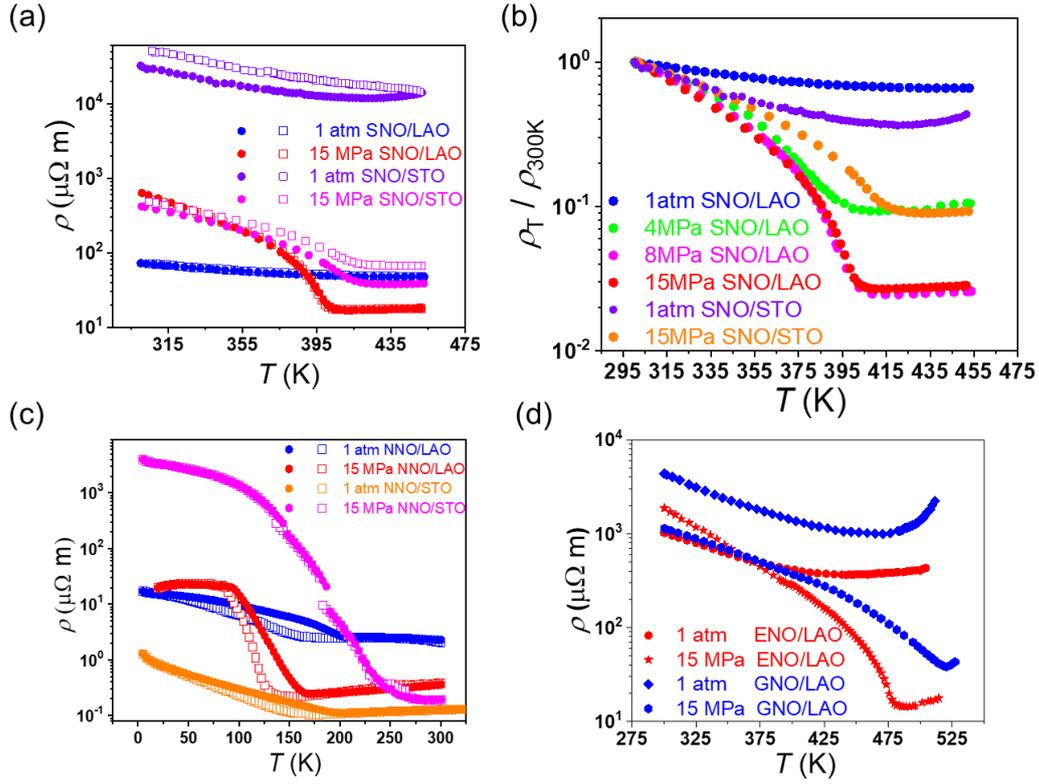

**Fig. 3.** (a) Temperature dependence of the resistivity for $SmNiO_3/LaAlO_3$ and $SmNiO_3/SrTiO_3$ under 1atm and 15MPa. (b) Temperature dependence of $\rho_T/\rho_{300K}$ for $SmNiO_3/LaAlO_3$ under various annealing oxygen pressure and for $SmNiO_3/SrTiO_3$ under 15MPa. (c) Temperature dependence of the resistivity for $NdNiO_3/LaAlO_3$ under 1atm and 15MPa and for $NdNiO_3/SrTiO_3$ under 15MPa. (d) Temperature dependence of resistivity for $EuNiO_3/LaAlO_3$ and $GdNiO_3/LaAlO_3$ under 1 atm and 15MPa. The solid symbols represent the heating processes, while the hollow symbols represent the cooling processes.

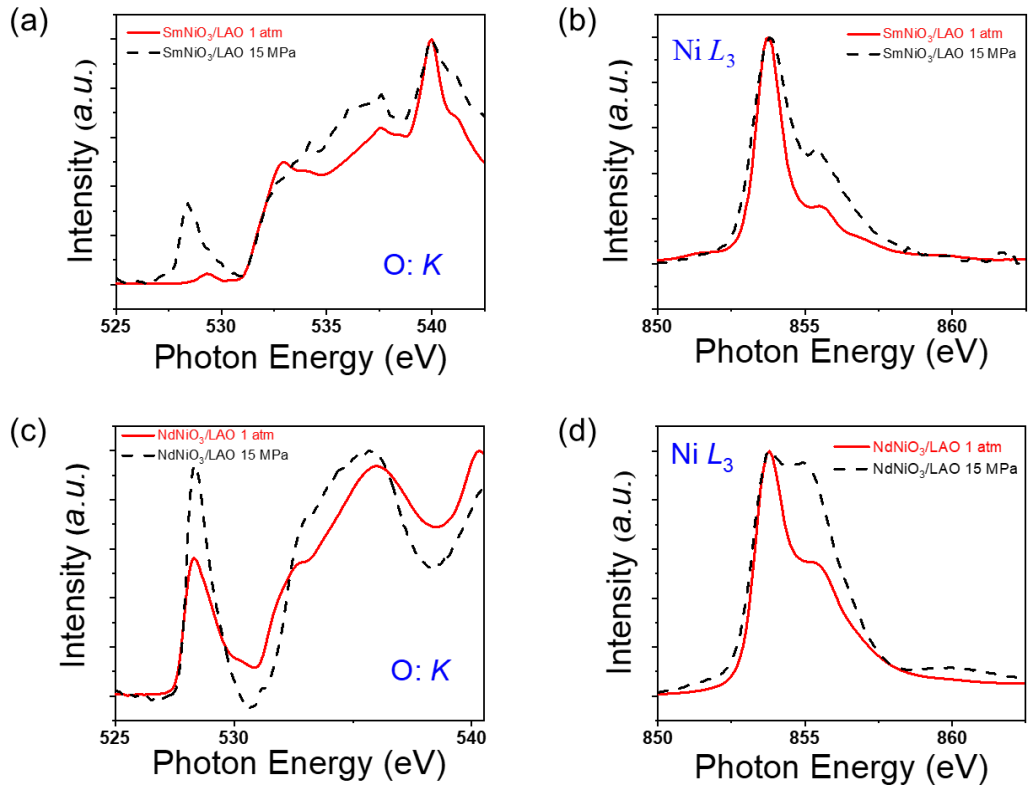

**Fig. 4.** Near edge X-ray absorption fine structure (NEXAFS) analysis of (a) O-K edge and (b) Ni-$L_3$ edge of SNO/LAO and (c) O-K edge and (d) Ni-$L_3$ edge of NNO/LAO with normal pressure and 15MPa. The data for SNO/LAO and NNO/LAO under 15MPa are from the reference [44].

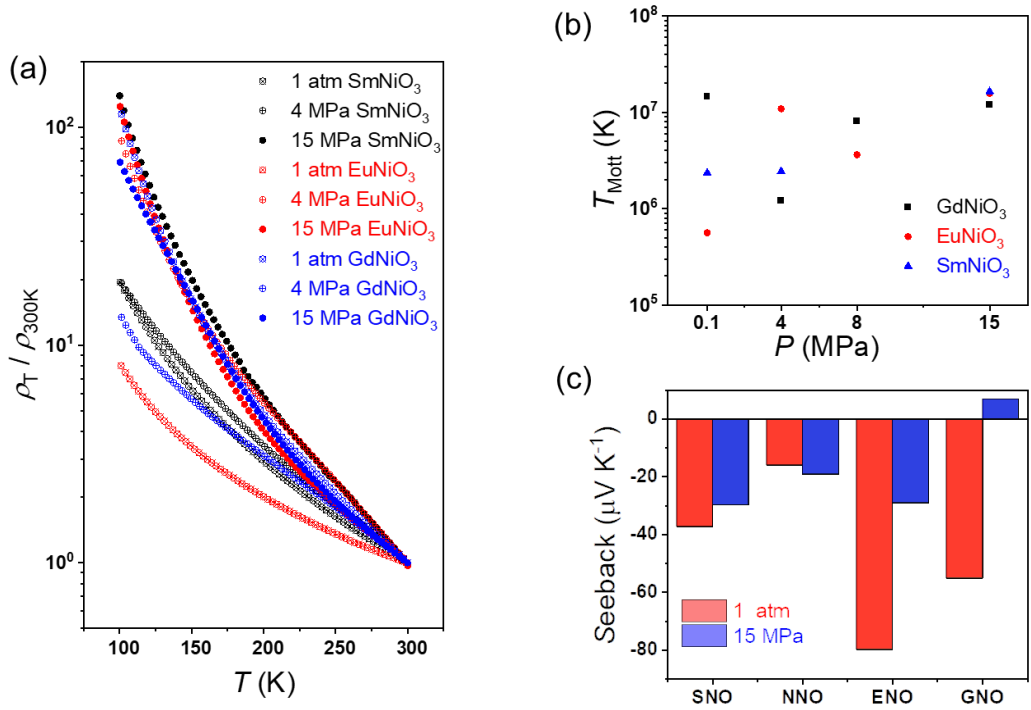

**Fig. 5.** (a) Illustration of $\rho_T/\rho_{300K}$ versus Temperature for SNO, ENO and GNO grown on LAO under various oxygen pressure. And the data for GNO under 15MPa are from the reference [53] (b) Oxygen pressure dependence of $T_{Mott}$ for SNO, ENO and GNO deposited on LAO. (c) Seebeck coefficients at room temperature under 1atm and 15MPa for SNO, NNO, ENO and GNO grown on LAO.

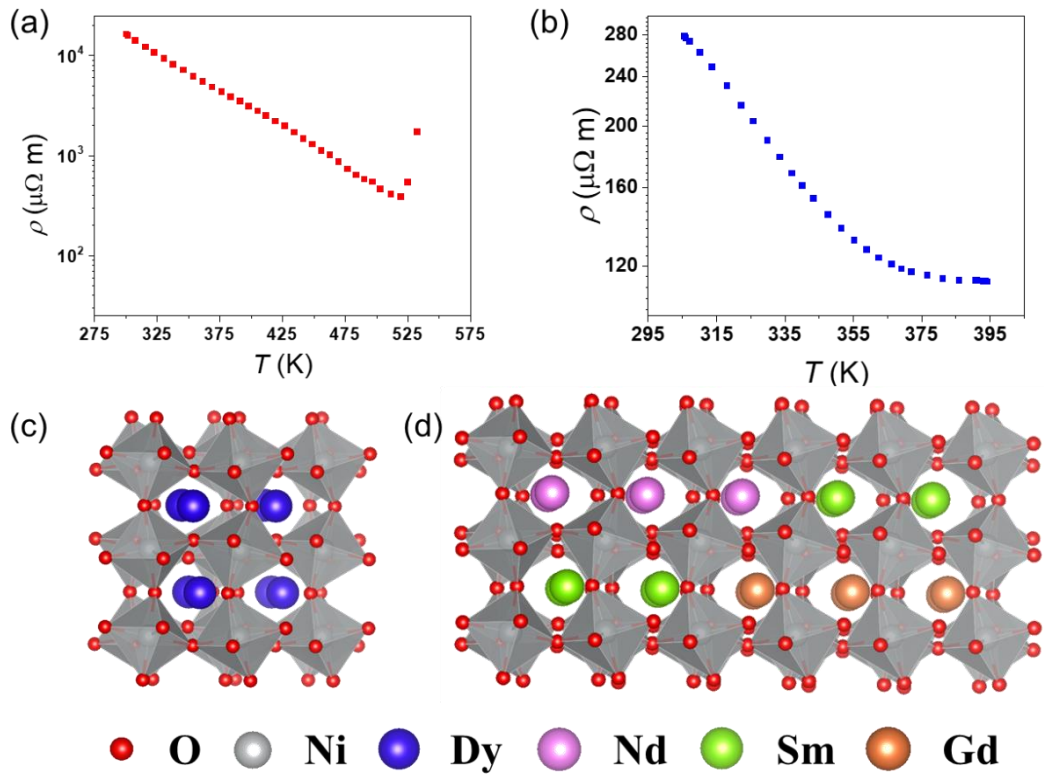

**Fig. 6.** Temperature dependence of the resistivity for (a) $DyNiO_3/LaAlO_3$ (001) and (b) $Nd_{0.3}Sm_{0.4}Gd_{0.3}NiO_3/LaAlO_3$ (001) under 15MPa. Schematic illustrations of the crystal structure for (c) $DyNiO_3$ and (d) $Nd_{0.3}Sm_{0.4}Gd_{0.3}NiO_3$, respectively.

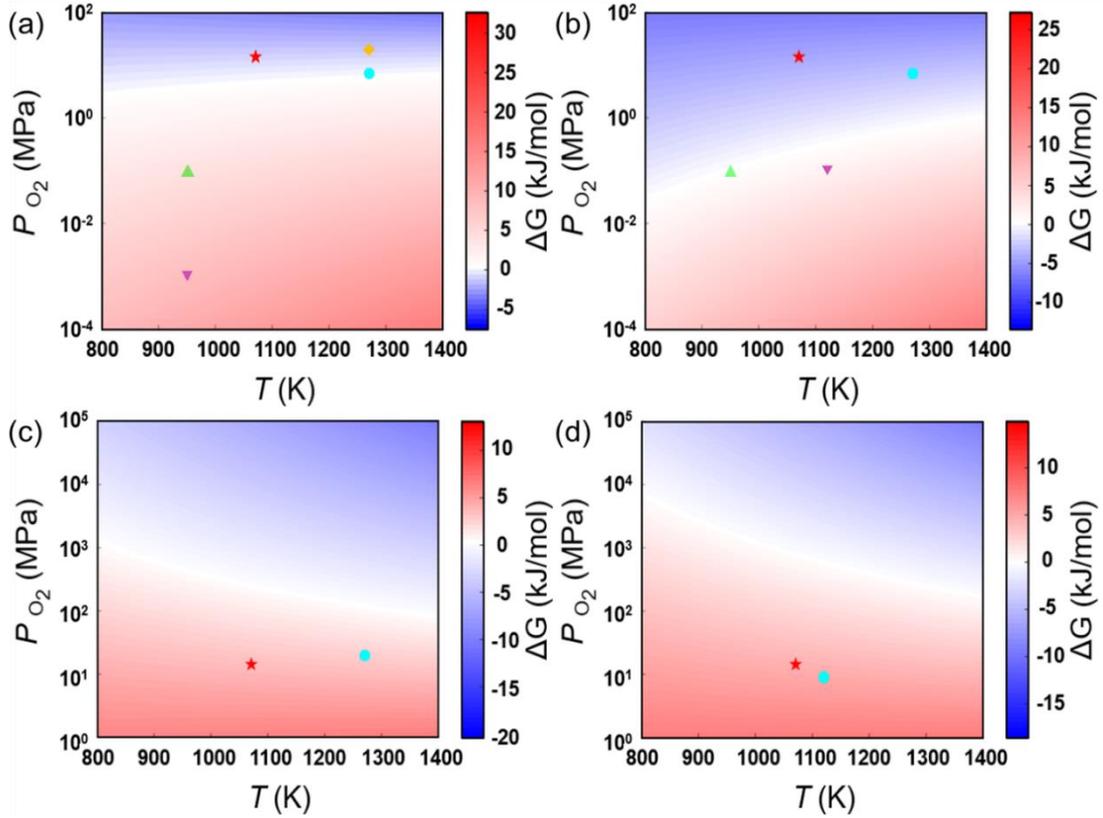

**Fig. 7.** Thermodynamic phase stability diagram for (a) SmNiO$_3$, Solid red star (1073K 15MPa) indicates conditions leading to SNO/LAO (001) epitaxial films in this work. Solid cyan circle (1273K 7MPa) and solid orange diamond (1273K 20MPa) indicate synthesis conditions of SmNiO$_3$ powder reported in the reference [2] and reference [62], respectively. Solid green upper triangular (953K 0.1MPa) and solid purple lower triangular (953K, 0.001MPa) indicate conditions of SmNiO$_3$ epitaxial films reported in the reference [34] and reference [63], respectively. (b) NdNiO$_3$, Solid red star (1073K 15MPa) indicates conditions leading to NNO/LAO (001) epitaxial films in this work. Solid cyan circle (1273K 7MPa) indicates conditions of NdNiO$_3$ powder reported in the reference [2]. Solid green upper triangular (953K 0.1MPa) and solid purple lower triangular (1123K 0.1MPa) suggest conditions of NNO/LAO (001) and NNO/LAO (110) epitaxial films reported in the reference [34] and reference [64], respectively. (c) EuNiO$_3$, Solid red star (1073K 15MPa) indicates conditions leading to ENO/LAO (001) epitaxial films in this work. Solid cyan circle (1273K 20MPa) represents condition for EuNiO$_3$ powder in the reference [65]. (d) GdNiO$_3$, Solid red star (1073K 15MPa) indicates conditions leading to GNO/LAO (001) epitaxial films in this work. Solid cyan circle (1123K 9MPa) indicates conditions leading to powder GdNiO$_3$ [55].

# Revealing the role of interfacial heterogeneous nucleation in metastable thin film growth of rare-earth nickelates electronic transition materials

*Fengbo Yan[1][†], Zhishan Mi[1,2][†], Jinhao Chen[1], Haiyang Hu[1], Lei Gao[1,3]\*, Jiaou Wang[4]\*, Nuofu Chen[5], Yong Jiang[1], Lijie Qiao[1,3] and Jikun Chen[1]\**

[1]Beijing Advanced Innovation Center for Materials Genome Engineering, School of Materials Science and Engineering, University of Science and Technology Beijing, Beijing 100083, China
[2]China Iron & Steel Research Institute Group, Material Digital R&D Center, Beijing 100081, China
[3]Corrosion and Protection Center, University of Science and Technology Beijing, Beijing 100083, China
[4]Beijing Synchrotron Radiation Facility, Institute of High Energy Physics, Chinese Academy of Sciences, Beijing 100049, China
[5]School of Renewable Energy, North China Electric Power University, Beijing 102206, China

Correspondence: Prof. Lei Gao (gaolei@ustb.edu.cn), Prof. Jiaou Wang (wangjo@ihep.ac.cn), and Prof. Jikun Chen (jikunchen@ustb.edu.cn).
[†]Fengbo Yan and Zhishan Mi contributed equally to this work.

**Section 1: XRD patterns at a broader scanning range for SNO/LAO, NNO/LAO, ENO/LAO and GNO/LAO**

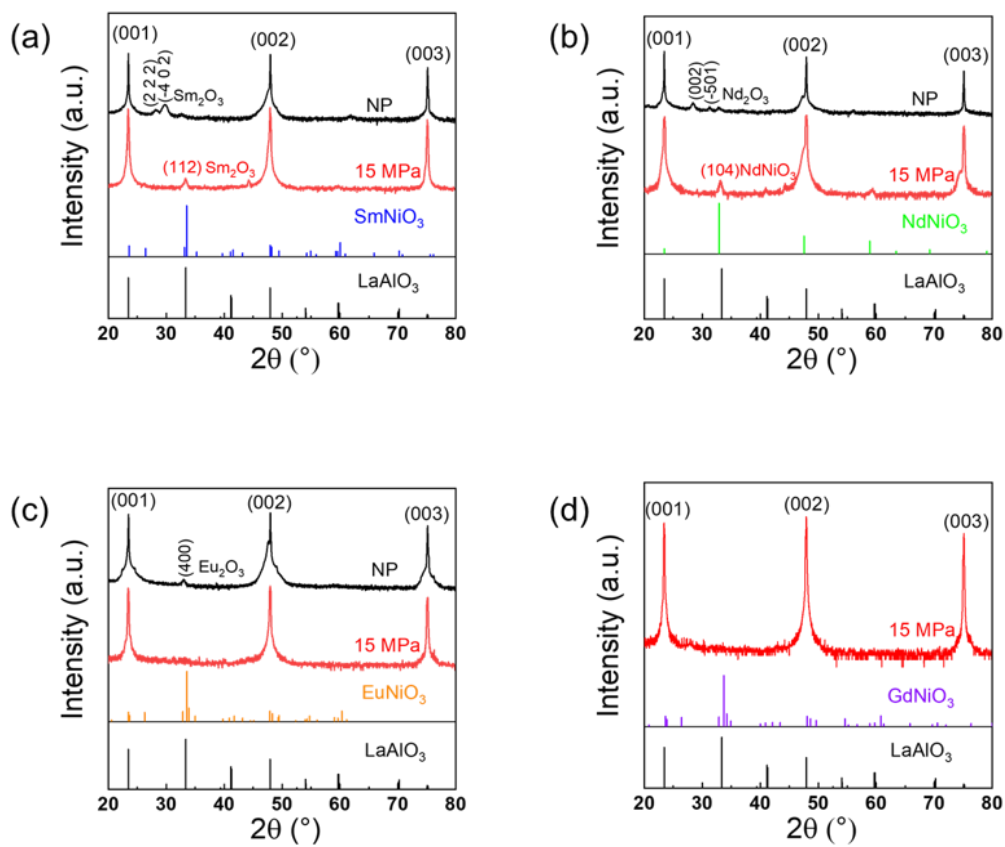

**Fig. S1.** X-ray diffraction patterns (θ-2θ scans) of as-grown (a) SmNiO$_3$/LaAlO$_3$(001) (b) NdNiO$_3$/LaAlO$_3$(001) (c) EuNiO$_3$/LaAlO$_3$(001) and (d) GdNiO$_3$/LaAlO$_3$(001) thin films from 20°-80° under various annealing oxygen pressures.

**Section 2: Calculation of lattice strain of the a-axis and b-axis**

LaAlO$_3$ has rhombohedral structure, which is defined as pseudo-cubic symmetry in order to finish the calculation. As for SrTiO$_3$, it is cubic structure. Therefore, we adopted the LaAlO$_3$ and SrTiO$_3$ as cubic structure with relaxed lattice constants of $a_0$=3.82 Å and $a_0$=3.96 Å, respectively. The orientation relationships between the rare-earth thin films and substrates follow (001)$_{film}$ // (001)$_{substrate}$, [100]$_{film}$ // [110]$_{substrate}$. According to this orientation relationship we built the interface supercells and set the reference lattice constants of interface structures as $b_0$=5.40 Å (LaAlO$_3$ interface) and $b_0$=5.60 Å (SrTiO$_3$ interface), respectively. The optimized lattice constants of rare-earth nickelates are listed in the following Table S1. And we calculated the lattice strain of a-axis and b-axis between rare-earth thin films and substrates using $\varepsilon_a$=(a-$b_0$)/$b_0$, $\varepsilon_b$=(b-$b_0$)/$b_0$. The obtained results are in the manuscript Table 1.

**Table S1**

The optimized lattice constants of rare-earth nickelates in a-axis and b-axis (unit: Å)

|  | a | b |
|---|---|---|
| SmNiO$_3$ | 5.28753 | 5.44663 |
| NdNiO$_3$ | 5.32534 | 5.46882 |
| EuNiO$_3$ | 5.33896 | 5.35262 |
| GdNiO$_3$ | 5.24442 | 5.51225 |

**Section 3: Interfacial charge transfer for *Re*NiO$_3$ grown on LAO(001) and STO(001) substrates**

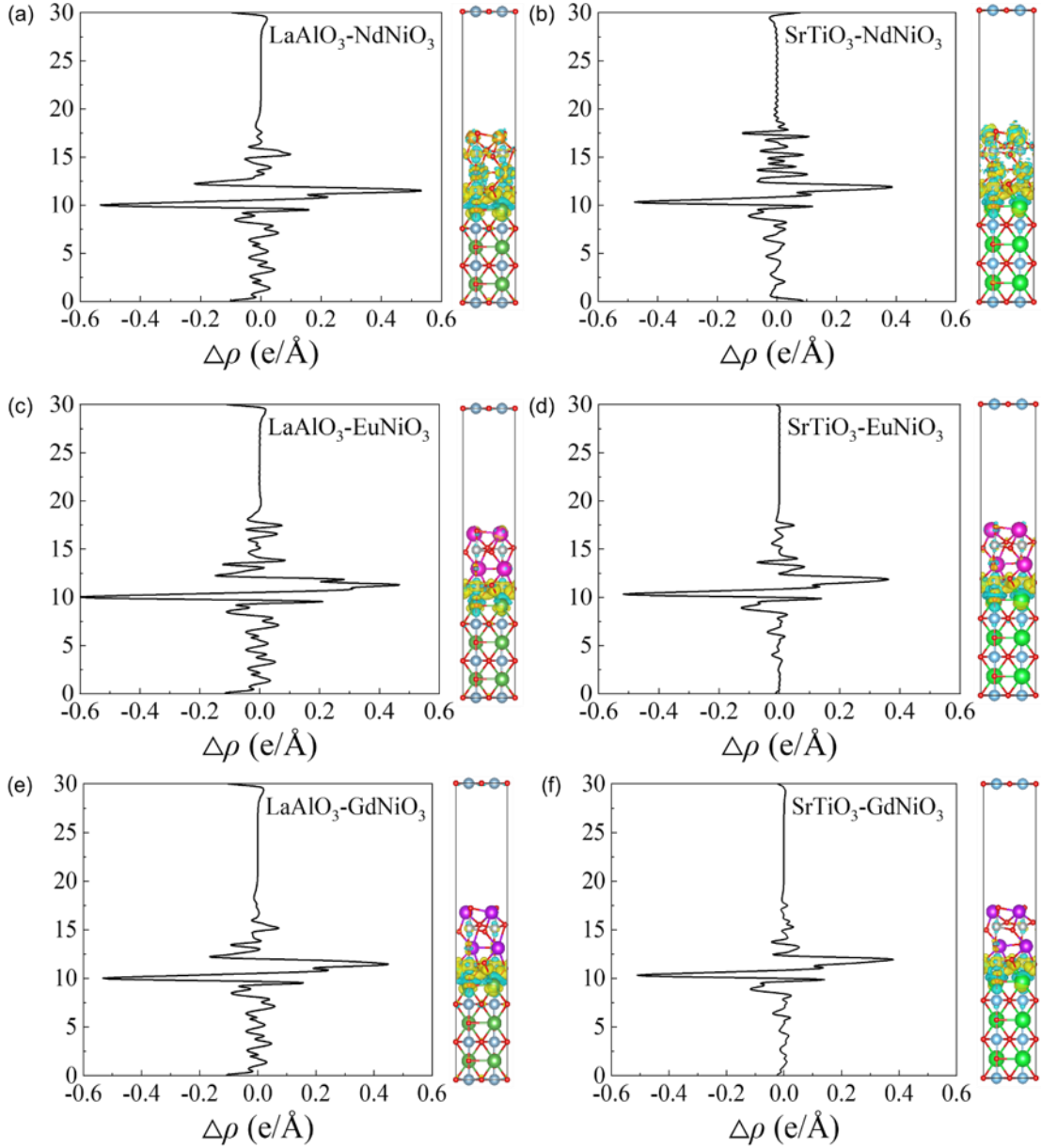

**Fig. S2.** The interfacial charge transfer of (a) LaAlO$_3$-NdNiO$_3$ interface (b) SrTiO$_3$-NdNiO$_3$ interface (c) LaAlO$_3$-EuNiO$_3$ interface (d) SrTiO$_3$-EuNiO$_3$ interface (e) LaAlO$_3$-GdNiO$_3$ interface (f) SrTiO$_3$-GdNiO$_3$ interface.

## Section 4: Details about Mott variable range hopping model fitting low temperature *R-T* curves of rare-earth nickelates

In this part, the process of how to use the Mott variable range hopping (VRH) [1] model to fit the low temperature *R-T* curves of rare-earth nickelates will be illustrated, from which we can obtain electrical parameters to describe the conduction mechanism at low temperature. The correlated formula [2] is given by

$$\rho(T) = \rho_0 \exp\left(\frac{T_0}{T}\right)^p \tag{S1}$$

where $\rho_0$ is the prefactor, $T_0$ is the characteristic temperature which represents the degree of disorder in the film [3], and *p* is the exponent depending on the conduction mechanism which is related to the density of states (*DOS*) near Fermi energy ($E_F$). In the light of Mott [4], when the *DOS* is constant, $p=1/(1+D)$ where D is the dimensionality of the conduction system, indicating $p=1/4$ in three dimensions. Localized charge carriers near the Fermi level can hop from one localized state to another due to thermal fluctuation or an electrical field provided by external factors. Fig. S4 shows the plots of ln *ρ(T)* versus $T^{-0.25}$ from 100K to a certain temperature decided by two temperature values, metal-insulator transition temperature and 300K for *Re*NiO$_3$ (*Re*=Sm, Eu, Gd) thin films grown on LaAlO$_3$ (001) substrates annealing at various oxygen pressures.

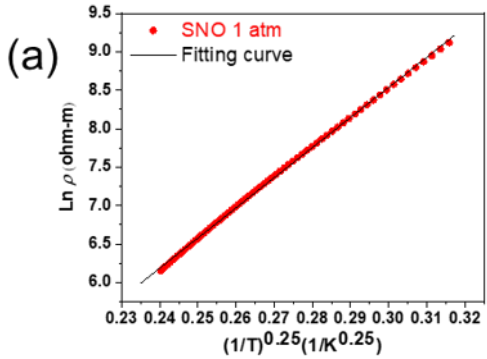
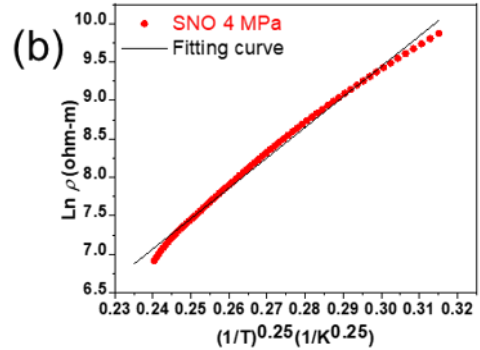
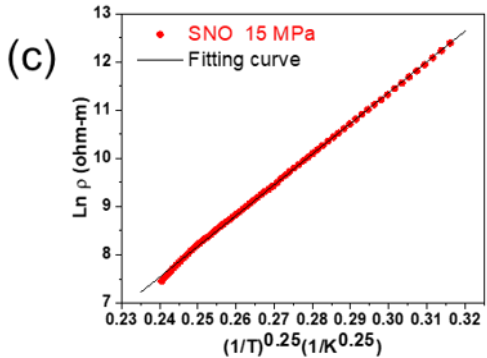

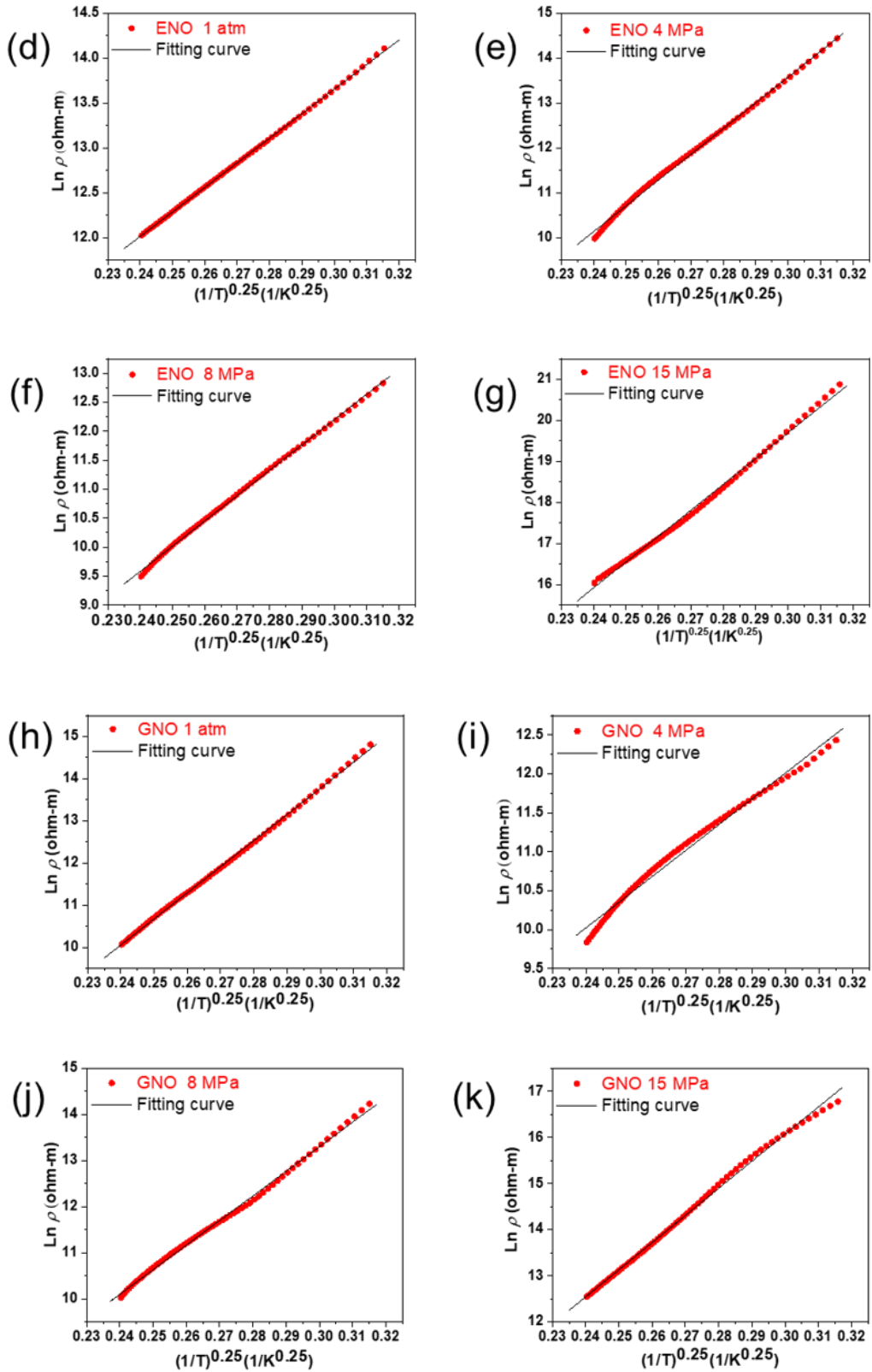

**Fig. S3.** Plots of Ln ($\rho_T$)-$T^{-0.25}$ for (a) SmNiO$_3$ 0.1 MPa (b) SmNiO$_3$ 4 MPa (c) SmNiO$_3$ 15 MPa (d) EuNiO$_3$ 0.1 MPa (e) EuNiO$_3$ 4 MPa (f) EuNiO$_3$ 8 MPa (g) EuNiO$_3$ 15 MPa (h) GdNiO$_3$ 0.1 MPa (i) GdNiO$_3$ 4 MPa (j) GdNiO$_3$ 8 MPa (k) GdNiO$_3$ 15 MPa (data from reference [5]) grown on LaAlO$_3$(001) substrates. Black lines are the linear fitting curves.

According to eqn (S1), when $p=1/4$, the corresponding $T_0$ ($T_{Mott}$) for each condition can be obtained from the slope of the fitting curve in Fig. S4, which is listed in the Table S2. On the other hand, according to the reference [3], the localization length can be calculated by

$$T_0 \equiv T_{Mott} = \frac{18}{k_B N(E_F) a^3} \tag{S2}$$

where $N(E_F)$ is the *DOS* near the $E_F$ and $a$ is the localization length. We take $N(E_F)$ =1.5×10$^{18}$ eV$^{-1}$cm$^{-3}$ [4]. The calculation results of localization length $a$ are also tabulated in the Table S2. The room temperature ($T$=300 K) optimum average hopping energy ($W_h$) and hopping distance ($R_h$) for Mott-VRH model are estimated using the following expressions

$$R_h = \left(\frac{9a}{8\pi k_B T N(E_F)}\right)^{1/4} \tag{S3}$$

$$W_h = \frac{3}{4\pi R_h^3 N(E_F)} \tag{S4}$$

In order to verify the validity of the fits for Mott VRH in *Re*NiO$_3$ thin films, it is necessary to calculate the values of $R_h/a$ and $W_h/k_B T$ listed in the Table S2 as well. Obviously, the two criteria (1) $R_h/a>1$ and (2) $W_h/k_B T>1$ are satisfied, which proves the consistency of the fits. As shown in Table S2, $T_{Mott}$ and localization length $a$ are found to be different for various annealing oxygen pressures, which means the annealing oxygen pressure has an influence on the film formation process. The results show a trend with an increase in $T_{Mott}$ as ρ of the film increases, which signifies films with high resistivity display a stronger localization behavior, which can be concluded from the Fig. S4 and Table S2. This trend has also been reported in other oxides [6,7].

**Table S2**
Parameters of Mott-VRH hopping mechanism in the *Re*NiO$_3$ (*Re*=Sm, Eu, Gd) thins films at various oxygen pressures.

| Sample | $T_{Mott}$(K) | $a$(nm) | $T_{max}$(K) | $R_{hmin}$(nm) | $R_h/a$ | $W_h$(eV) | $W_h/k_B T$ |
|---|---|---|---|---|---|---|---|
| SNO 0.1 MPa | 2.35E+06 | 3.90 | 300 | 13.77 | 3.53 | 0.0609 | 2.36 |
| SNO 4 MPa | 2.44E+06 | 3.85 | 300 | 13.73 | 3.57 | 0.0615 | 2.38 |
| SNO 15 MPa | 1.65E+07 | 2.04 | 300 | 11.71 | 5.75 | 0.0991 | 3.83 |
| ENO 0.1 MPa | 5.60E+05 | 6.29 | 300 | 15.52 | 2.47 | 0.0425 | 1.64 |
| ENO 4 MPa | 1.09E+07 | 2.34 | 300 | 12.12 | 5.19 | 0.0894 | 3.46 |
| ENO 8 MPa | 3.63E+06 | 3.37 | 300 | 13.28 | 3.94 | 0.0679 | 2.63 |
| ENO 15 MPa | 1.58E+07 | 2.07 | 300 | 11.75 | 5.69 | 0.0981 | 3.79 |
| GNO 0.1 MPa | 1.47E+07 | 2.12 | 300 | 11.82 | 5.59 | 0.0963 | 3.73 |
| GNO 4 MPa | 1.22E+06 | 4.85 | 300 | 14.55 | 3.00 | 0.0517 | 2.00 |
| GNO 8 MPa | 8.08E+06 | 2.58 | 300 | 12.43 | 4.81 | 0.0829 | 3.21 |
| GNO 15 MPa | 1.20E+07 | 2.26 | 300 | 12.02 | 5.31 | 0.0915 | 3.54 |

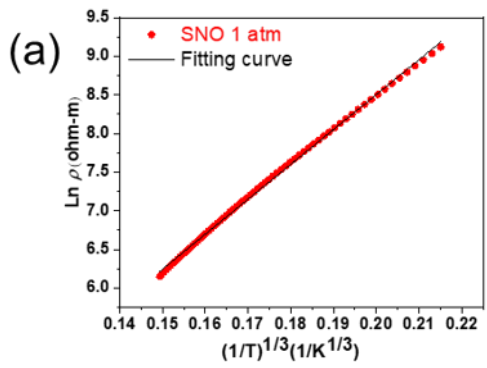
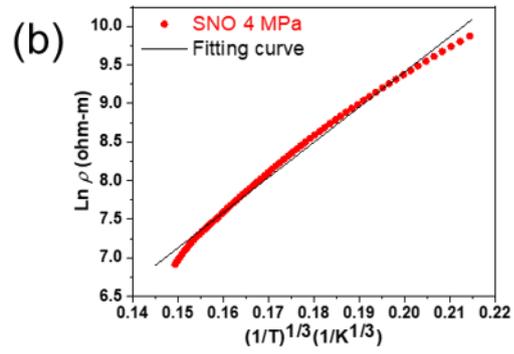
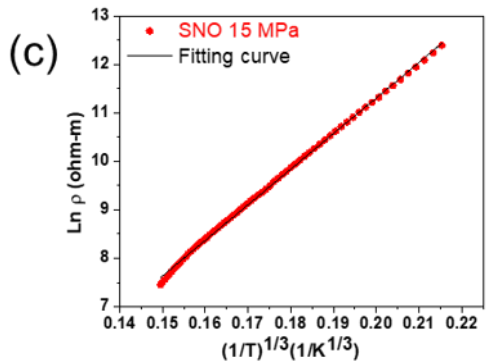
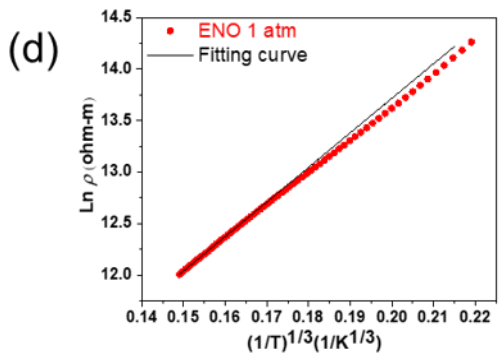
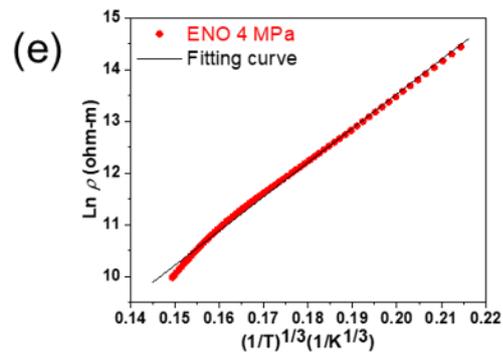
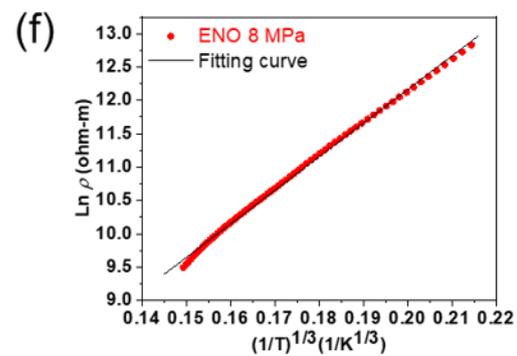
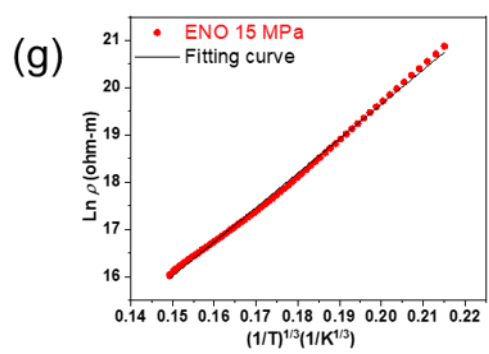

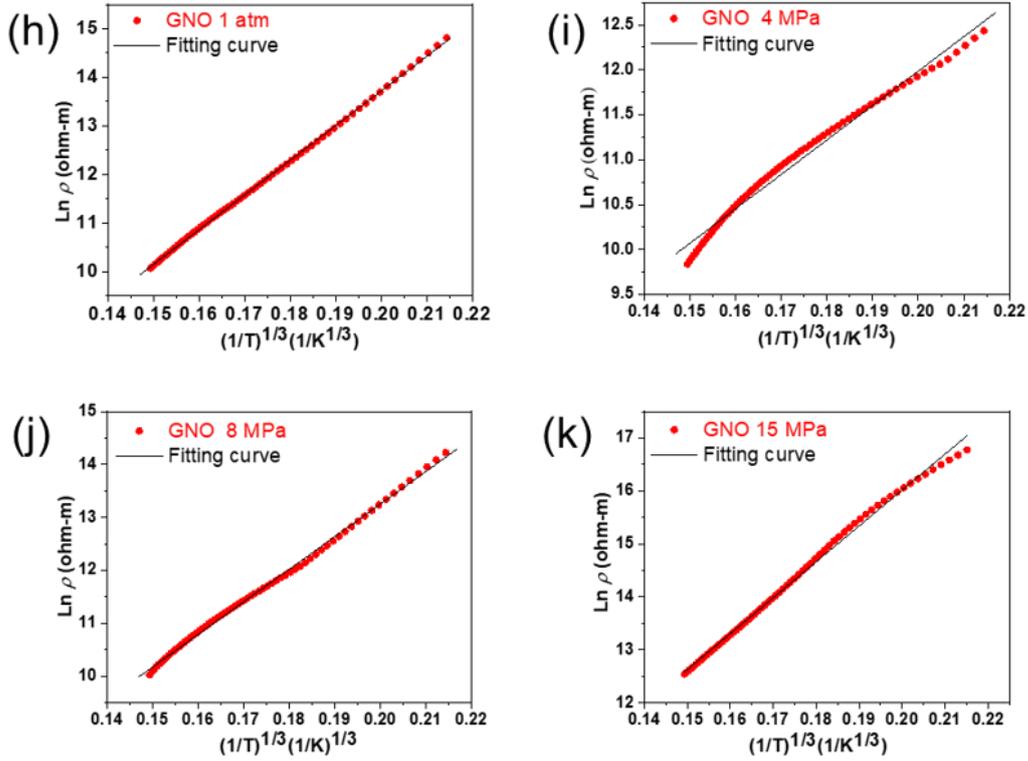

**Fig. S4.** Plots of Ln ($\rho_T$)-$T^{-1/3}$ for (a) SmNiO$_3$ 0.1 MPa (b) SmNiO$_3$ 4 MPa (c) SmNiO$_3$ 15 MPa (d) EuNiO$_3$ 0.1 MPa (e) EuNiO$_3$ 4 MPa (f) EuNiO$_3$ 8 MPa (g) EuNiO$_3$ 15 MPa (h) GdNiO$_3$ 0.1 MPa (i) GdNiO$_3$ 4 MPa (j) GdNiO$_3$ 8 MPa (k) GdNiO$_3$ 15 MPa (data from reference [5]) grown on LaAlO$_3$(001) substrates. Black lines are the linear fitting curves.

Considering the thickness of as-grown $Re$NiO$_3$ thin films (about 10 nm), we also fit the ln($\rho_T$)-$T$ tendencies via two-dimensional carrier transport model. For the situation of two dimensions, $p=1/3$. Using similar method, the fitting plots of ln($\rho_T$)-$T^{-1/3}$ for various $Re$NiO$_3$ can be obtained in Fig. S5. According to the reference [3], the localization can be obtained by

$$T_0 \equiv T_{Mott} = \frac{13}{k_B N(E_F) a^2} \quad (S5)$$

where $N(E_F)$ is the *DOS* near the $E_F$ and $a$ is the localization length. The calculation results are listed in Table S3.

In order to verify the correctness of the fits with the Mott VRH for $Re$NiO$_3$, we calculate the hopping distance $R_h$ by formula (S6) and the average hopping energy $W_h$ by formula (S7) to confirm whether the two conditions $R_h/a>1$ and $W_h/(k_B T)>1$ are satisfied. The results are listed in the table S3 as well.

$$R_h = \left(\frac{9a}{8\pi k_B T N(E_F)}\right)^{1/3} \quad (S6)$$

$$W_h = \frac{3}{4\pi R_h^2 N(E_F)} \quad (S7)$$

**Table S3**
Parameters of Mott-VRH hopping mechanism in the $ReNiO_3$ ($Re$=Sm, Eu, Gd) thins films at various oxygen pressure.

| Sample | $T_{Mott}$(K) | $a$(nm) | $T_{max}$(K) | $R_{hmin}$(nm) | $R_h/a$ | $W_h$(eV) | $W_h/k_BT$ |
|---|---|---|---|---|---|---|---|
| SNO 0.1 MPa | 9.21E+04 | 0.010 | 300 | 0.021 | 2.04 | 0.0351 | 1.36 |
| SNO 4 MPa | 9.46E+04 | 0.010 | 300 | 0.021 | 2.06 | 0.0355 | 1.37 |
| SNO 15 MPa | 3.97E+05 | 0.005 | 300 | 0.017 | 3.32 | 0.0572 | 2.21 |
| ENO 0.1 MPa | 3.77E+04 | 0.016 | 300 | 0.025 | 1.51 | 0.0261 | 1.01 |
| ENO 4 MPa | 2.91E+05 | 0.006 | 300 | 0.018 | 2.99 | 0.0515 | 1.99 |
| ENO 8 MPa | 1.28E+05 | 0.009 | 300 | 0.020 | 2.27 | 0.0392 | 1.52 |
| ENO 15 MPa | 3.86E+05 | 0.005 | 300 | 0.017 | 3.28 | 0.0567 | 2.19 |
| GNO 0.1 MPa | 3.66E+05 | 0.013 | 300 | 0.017 | 3.22 | 0.0556 | 2.15 |
| GNO 4 MPa | 5.62E+04 | 0.007 | 300 | 0.023 | 1.72 | 0.0298 | 1.15 |
| GNO 8 MPa | 2.33E+05 | 0.006 | 300 | 0.018 | 2.78 | 0.0479 | 1.85 |
| GNO 15 MPa | 3.12E+05 | 2.26 | 300 | 0.017 | 3.06 | 0.0528 | 2.14 |

We can notice that similar results are in two dimensions as in three dimensions, which shows the two dimensions Mott VRH model is also satisfied generally.

**Section 5: Estimation of parameters of $h$ and $s$ in molar Gibbs free energy of formation of bulk $ReNiO_3$**

According to the reference [8], the molar Gibbs free energy of formation of bulk $ReNiO_3$ can be estimated from the following:

$$\Delta G = \Delta H_{LNO,1000k} - T\Delta S_{LNO,1000K} + (h - sT)(r(Re^{3+}) - r(La^{3+})) - (1/4)RT\ln(P/p) \tag{S8}$$

where LNO stands for LaNiO$_3$, $T$ is the absolute temperature, $R$ is the ideal gas constant, the $P$ represents oxygen pressure with MPa unit, $p$ represents the standard atmospheric pressure (0.1 MPa) and $r(Re^{3+})$ and $r(La^{3+})$ are the radii of $Re^{3+}$ and $La^{3+}$, respectively. According to the reference [9], the enthalpy and entropy changes for the formation of LaNiO$_3$ from oxides and O$_2$ at 1000K are $\Delta H_{LNO,1000K}$=-46.07 kJmol$^{-1}$ and $\Delta S_{LNO,1000K}$=-26.4×10$^{-3}$ kJK$^{-1}$mol$^{-1}$. $h$ and $s$ represent the trend in enthalpy and entropy with variation of ionic radii. They can be estimated from the following process. As mentioned in the reference [8], the value of $h$ is calculated using a statistically weighted averaged of linear slope from $\Delta G$ vs. r(Re$^{3+}$) for $Re$MnO$_3$, $Re$FeO$_3$ and $Re$CoO$_3$, which has been obtained that $h$ = -207 kJmol$^{-1}$Å$^{-1}$. As a typical rare-earth nickelate perovskite, the synthesis conditions of SmNiO$_3$ have been widely investigated. It is accepted that the molar Gibbs free energy of formation of bulk SmNiO$_3$ is zero at the lowest pressure for high temperature synthesis of SmNiO$_3$, and the lowest pressure for successful high temperature synthesis of SmNiO$_3$ is reported by Escote *et al.* (1273 K, 70 bar) [10]. This condition was further used to estimate the value of $s$. On the other hand, the radii of rare-earth elements are obtained from the reference [11]. Based on these conditions, the value of $s$ is estimated as -0.00744 kJK$^{-1}$mol$^{-1}$Å$^{-1}$. It is clear the value of $s$ is much smaller than the value of $h$, indicating the assumption is effective.